\documentclass[twocolumn,showpacs,preprintnumbers,amsmath,amssymb]{revtex4}

\usepackage{graphicx}

\begin{document}

\title{Electron concentration effects on the Shastry-Sutherland phase stability in Ce$_{2-x}$Pd$_{2+y}$In$_{1-z}$ solid solutions}

\author{J.G. Sereni$^1$, M. Giovannini$^2$, M. G\'omez Berisso$^1$, A. Saccone$^2$}
\address{$^1$Div. Bajas Temperaturas, Centro At\'omico Bariloche (CNEA) and Conicet, 8400 S.C. Bariloche, Argentina\\
CNR-SPIN, Dipartimento di Chimica e Chimica Industriale,
Universit\'a di Genova, I-16146 Genova, Italy}

\date{\today}

\begin{abstract}

{The stability of a Shastry-Sutherland ShSu phase as a function of
electron concentration is investigated through the field
dependence of thermal and magnetic properties of the solid
solution Ce$_{2-x}$Pd$_{2+y}$In$_{1-z}$ in the antiferromagnetic
branch. In these alloys the electronic (holes) variation is
realized by increasing $Pd$ concentration. The AF transition $T_M$
decreases from 3.5\,K to 2.8\,K as $Pd$ concentration increases
from $y=0.2$ to $y=0.4$. By applying magnetic field, the ShSu
phase is suppressed once the field induced ferromagnetic
polarization takes over at a critical field $B_{cr}$ which
increases with $Pd$ content. A detailed analysis around the
critical point reveals a structure in the maximum of the $\partial
M/\partial B$ derivative, which is related with incipient steps in
the magnetization $M(B)$ as predicted by the theory for the ShSu
lattice. The crossing of $M(B)$ isotherms, observed in ShSu
prototype compounds, is also analyzed. The effect of $In$
substitution by $Pd$ is interpreted as an increase of the number
of 'holes' in the conduction band and results in a unique
parameter able to describe the variation of the magnetic
properties along the studied range of concentration.}

\end{abstract}

 \maketitle

\section{Introduction}

Magnetic frustration is an attracting topic of physics since it
may drive the systems to new and exotic phases \cite{Lacriox}.
Frustration is mostly originated in peculiar geometrical
conditions which impede the development of long range order. One
of the simplest examples of magnetic frustration is a triangular
coordination of moments interacting antiferromagnetically AF among
them. Since a canonical AF minimum of energy cannot be reached,
the system accesses to alternative minima where other types of
order parameters may develop.

In the presence of AF interactions, Shastry-Sutherland ShSu
lattices \cite{Shastry} present those characteristics because
neighboring magnetic atoms are disposed in triangular
arrangements. In the model compound SrCu$_2$(BO$_3$)$_2$
\cite{Kageyama}, magnetic Cu$^{2+}$ atoms have one
nearest-neighbor $nn$ and four next-nearest-neighbors $nnn$ on the
same plane, all of them coupled AF by respective $J$ and $J'$
exchanges. Since the effective interactions are $J > J'$, $nn$
atoms form a network of $J$ mediated orthogonal dimers, being the
interaction between dimers mediated by $J'$. As a result, the
magnetic structure can be described as a quasi 2D lattice of
orthogonal dimers \cite{Miyahara}.

Two significant features were observed in the field $B$ dependent
magnetization $M$ curves of SrCu$_2$(BO$_3$)$_2$ \cite{Kageyama}:
i) a crossing of the magnetization isotherms at low temperature
($T<4$\,K) and ii) two small steps in $M(B)$ at fractional values
(c.f. 1/4 and 1/8) of the saturation moment $M_{sat}$. In this
compound, those steps are observed at a quite high applied
magnetic field $B\approx 20$\,T \cite{Kageyama}. This so-called
'quantized magnetization' scenario, with M=1/2, 1/4, 1/8.... of
M$_{sat}$ is explained by the theory \cite{Miyahara} as due to
successive commensurtate shells of 2D squares involving integer
number of dimers. This model considers the geometrical
distribution of the magnetic entities (i.e. dimers) as
topologically equivalent to the 2D square lattice of the
Heisenberg model. A simple series expansion according to $N=2^n$,
where $n$ is an integer number $n=1,2,....$, accounts for the
amount of dimers $N$ involved in the successive shells as the
magnetic field increases. As a consequence of the $J$-AF
character, the four fold state of the dimers splits into a singlet
ground state GS and an excited triplet \cite{Miyahara}.

Among intermetallic compounds, some heavy rare earth tetraborides
RB$_4$ \cite{Michimura} were found to show these features but at
lower magnetic field, e.g. 4T for ErB$_4$, with a clear step at
$1/2M_{sat}$. Similar scenario was invoked in Yb$_2$Pt$_2$Pb
\cite{Kim} which orders AF at $T_M\approx 2$\,K. Within the
Mo$_2$B$_2$Fe type structure there is a large family of compounds
with the formula R$_2$T$_2$X, being T a transition metal and X a
p-metal. In this structure, the magnetic atoms 'R' are disposed in
a unique crystalline lattice. Particularly, those compounds with T
= Pd and X= Sn \cite{Fourgeot} or In \cite{Mauro98} are of
interest concerning the ShSu phase formation because of the
triangular coordination of the 'R' magnetic atoms. This quasi 2D
structure can be described as successive 'T+X' (at $z=0$) and 'R'
(at $z=1/2$) layers \cite{Peron}, with the 'R' $nn$ and $nnn$
disposed with the same symmetry like in SrCu$_2$(BO$_3$)$_2$,
albeit at different relative distances.

A ShSu phase was also reported in Ce$_2$Pd$_2$Sn \cite{Ce2Pd2Sn}
within a limited range of temperature, i.e. between an AF
transition at $T_M=4.9$\,K and a ferromagnetic FM one of first
order type at $T_C=2.1$\,K. Since in this compound dimer's
formation is mediated by a $J$-FM exchange, the previously
mentioned triplet has the lowest energy. The ShSu phase is built
by a $J'$-AF exchange below $T_M=4.9$\,K, however it becomes
instable below $T_C=2.1$\'K when the interplane FM interaction
$J_{C}$ takes over inducing a 3D-FM ground state GS.

The ShSu phase is suppressed under a relatively low magnetic field
at a magnetic critical point: $B_{cr}=0.11$\,T and $T_{cr}=4.1$\,K
\cite{Ce2Pd2SnB}. Despite of the low $B_{cr}$ value, isothermal
$M(B)$ measurements show the features observed in the mentioned
ShSu model compounds, i.e. the crossing of magnetization isotherms
and an incipient step as a function of field at $M\approx 1/4
M_{sat}$.

The stability of the ShSu phase was also investigated under
structural pressure, by doping $Pd$ with smaller isoelectronic
$Ni$ atoms. The expected weakening of the Ce magnetic moment and
the increase of the Sommerfeld coefficient $\gamma$ occurs in
Ce$_2$(Pd$_{1-x}$Ni$_x$)$_2$Sn as the Kondo screening increases.
The upper transition temperature $T_M(Ni)$ decreases due to the
weakening of the $J$ exchange \cite{SCES2010}. On the contrary,
the lower transition $T_C$ to the FM phase increases with $x(Ni)$
as a consequence of the reduction of the temperature range of the
ShSu stability because of the weakening of the $J'$ coupling. Both
$J$ and $J'$ are overcome by the inter-Ce planes coupling in the
'c' crystalline direction $J_C$, which for $x=0.25$ practically
inhibits the ShSu phase formation.

Besides magnetic field and structural pressure, the third
complementary control parameter to be used to investigate this
ShSu phase stability is the chemical potential variation. In this
family of compounds, the significant range of solubility of the
2-2-1 ternary indides \cite{Mauro00} provides the possibility to
investigate the effect of the electronic concentration.
Particularly, within the region where $In$ sites can be occupied
by $Pd$ atoms, the number of 'holes' in the conduction band is
expected to increase driven by $Pd$ concentration. For such a
purpose, we have performed a systematic study of the low
temperature magnetic properties of Ce$_{2-x}$Pd$_{2+y}$In$_{1-z}$
alloys within a broad range of composition.

\begin{figure}
\begin{center}
\includegraphics[angle=0, width=0.55 \textwidth] {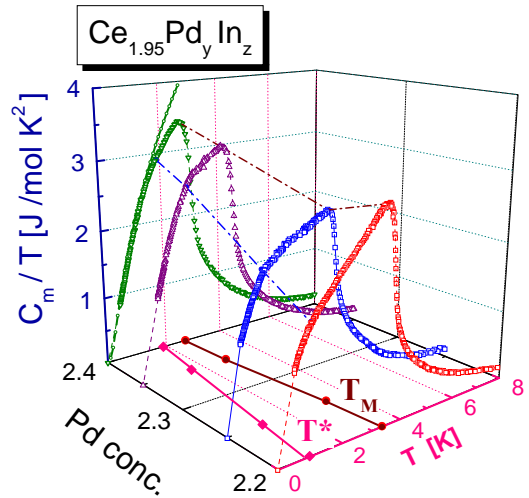}
\end{center}
\caption{(Color online) Thermal dependence of the magnetic
contribution to specific heat $C_m$ divided T for different $Pd$
concentrations. Dash-dot curves are guides to the eye indicating
$T_M$ and $T^*$ variations, which are also projected on the basal
plane. Continuous curve on the Ce$_{1.95}$Pd$_{2.4}$In$_{0.7}$
compound is the fit for $T\leq T^*$, see the text.} \label{F1}
\end{figure}

\section{Experimental details and results}

The metals used for sample preparation were palladium (foil, 99.95
mass \% purity, Chimet, Arezzo, Italy), cerium (bar, 99.99 mass \%
purity, NewMet Kock, Waltham Abey, UK) and indium (ingot, 99.999
\% mass, Johnson Matthey, London, UK). The samples, each with a
total weight of about 2 g, were prepared by weighing the proper
amounts of elements and then by argon arc melting the elements on
a water cooled copper hearth with a  tungsten electrode. To ensure
good homogeneity the buttons were turned over and remelted several
times. Weight losses after melting were always smaller than 0.5
mass \%. All the samples were then annealed at 750 °C for 10 days,
and finally quenched in cold water. Scanning electron microscopy
SEM supplied by Carl Zeiss SMT Ltd. Cambridge, England, and
electron probe micro-analysis EPMA based on energy-dispersive
X-ray spectroscopy were used to examine phase compositions. Smooth
surfaces of specimens for microscopic observation were prepared by
using SiC papers and diamond pastes down to 1\,$\mu$ grain size.
The compositional contrast was revealed in un-etched samples by
means of a backscattered electron detector BSE. For the
quantitative analysis an acceleration voltage of 20\,KV was
applied for 100 s, and a cobalt standard was used for calibration.
The X-ray intensities were corrected for ZAF effects using the
pure elements as standards. The composition values derived were
usually accurate to 1\,at\%. X-ray diffraction XRD was performed
on powder samples using the vertical diffractometer X-Pert MPD
(Philips, Almelo The Netherland) with Cu K  radiation.

Specific heat was measured using a standard heat pulse technique
in a semi-adiabatic He-3 calorimeter in the range between 0.5 and
20K, at zero and applied magnetic field up to 2T. DC-magnetization
measurements were carried out using a standard SQUID magnetometer
operating between 2 and 300K, and as a function of field up to 5T.
For AC-susceptibility measurements a lock-in amplifier was used
operating at 1.28KHz, with an excitation field of 1Oe on
compensated secondary coils in the range of 0.5 to 10K.

The Ce$_2$Pd$_2$In system presents a wide range of solid solution
\cite{Mauro00}, with a gap of solubility around the stoichiometric
composition. Such a gap splits the range of solubility into two
'branches' which show distinct magnetic structures, being the
Ce-rich branch FM and the Pd-rich one AF \cite{Mauro00}. The
studied alloys can be described by the general formula
Ce$_{2-x}$Pd$_{2+y}$In$_{1-z}$, with: $-0.1\leq x \leq 0.1$, $y =
x + z$ and $0.05 \leq z \leq 0.3$, with the richest $Pd$
composition placed at the edge of the chemical stability. Most of
the work was performed on the AF-branch, but one alloy of the
FM-branch (c.f. Ce$_{2.1}$Pd$_{1.95}$In$_{0.95}$) was also
investigated for comparison with the aim to extend the study of
the electron concentration effect.

Concerning the structural consequences of the $Pd/In$ substitution
in the alloys belonging to the $Pd$-rich AF-branch, one can
recognize from the compositional dependence of the lattice
parameters that the volume cell practically does not change. There
is, however, a slight decrease of the 'a' lattice parameter
accompanied by an increase of the 'c' parameter, which produces an
increase of the 'c/a' ratio. Such a variation is explained by the
fact that the excess of $Pd$ replacing larger In atoms occupies
two $4e$ sites which lie on the 'c' axis symmetrically placed
respect to the original $In-2a$ site \cite{Mauro00}. As a
consequence the Ce-Ce inter-planes distance is expected to
increase with $Pd$ content.

\begin{figure}
\begin{center}
\includegraphics[angle=0, width=0.5 \textwidth] {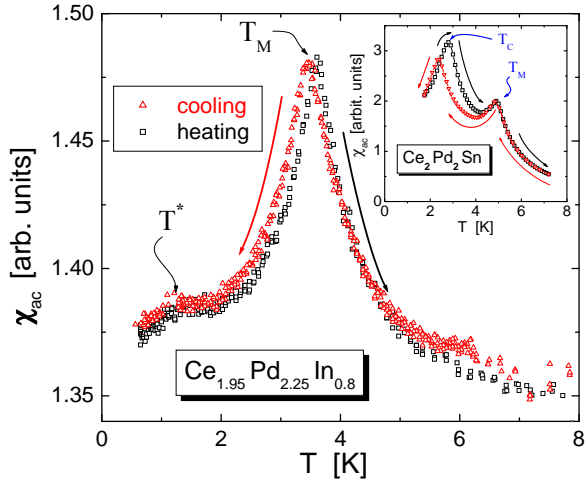}
\end{center}
\caption{(Color online) Low temperature ac-susceptibility of
Ce$_{1.95}$Pd$_{2.25}$In$_{0.8}$. Inset: comparison with
Ce$_2$Pd$_2$Sn after Ref.\cite{Ce2Pd2Sn}.} \label{F2}
\end{figure}

\subsubsection{Thermal Properties}

In Fig.~\ref{F1} we compare the thermal dependence of the magnetic
contribution to the specific heat for samples with concentration
$Pd(2+y)=$ 2.20, 2.25, 2.35 and 2.40, all of them belonging to the
Pd-rich AF 'branch'. The magnetic contribution $C_m(T)$ is
computed by subtracting the $C_P(T)$ of the non-magnetic isotypic
La compound to the measured value. In the figure one can see that
the transition $T_M$ decreases upon increasing $Pd$ concentration
from 3.5 to 2.8\,K (see the $T_M$ projection on the basal plane).
The maximum value of $C_m/T_M$ first decreases between $Pd(2+y)=$
2.20 and 2.25, and then increases monotonously up to
$Pd(2+y)=2.40$. This non monotonous behavior can be attributed to
different competing contributions which change differently with
$Pd$ content. The $C_m(T_M)$ transition of the $Pd(2+y)= 2.20$
shows a sharper and higher maximum than the other alloys, probably
due to its lower disorder due to its proximity to stoichiometry.
For higher $Pd$ concentration, the maximum
increases because of the $C_m(T_M)$ ratio with decreasing
$T_M(Pd)$ values.

Contrary to Ce$_2$Pd$_2$Sn, below $T_M$ there is no first order
transition (like $T_C$), but a shoulder at a temperature $T=T^*$
which increases with $Pd$ content. Such a temperature is traced as
the maximum curvature of $C_m(T)/T$. This weak change in the
thermal dependence of $C_m/T$ indicates that around that
temperature the modulated phase \cite{Mauro00} continuously
transforms into the FM GS.

\begin{figure}
\begin{center}
\includegraphics[angle=0, width=0.5 \textwidth] {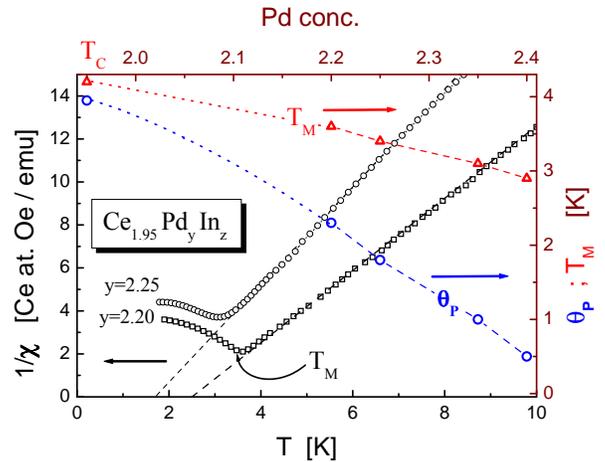}
\end{center}
\caption{(Color online) Low temperature inverse susceptibility
($B/M$, with $B=0.3T$) of two samples showing the paramagnetic
temperature extrapolation $\theta_P$ (left and lower axes). $Pd$
concentration dependence of $\theta_P$ and susceptibility cusp at
$T=T_M$ (right and upper axes), including the FM sample. Dot lines
are the extrapolations of $\theta_P$ and $T_M$ parameters from AF
samples (dashed lines) to those of the FM one.} \label{F3}
\end{figure}

In order to check this difference between indides and stanides
compounds, we have measured the ac-susceptibility $\chi_{ac}$ on
the $Pd(2+y)=2.25$ alloy, see Fig.~\ref{F2}. The $\chi_{ac}(T)$
maximum at $T= 3.6$\,K coincides with the corresponding $T_M$
value, but only a weak shoulder is detected at $T= 1.5$\,K in
coincidence with $C_m(T)/T$. This temperature dependence has to be
compared with that of Ce$_2$Pd$_2$Sn \cite{Ce2Pd2Sn} shown in the
inset. There, the first order character of the lower transition is
clearly evidenced by the thermal hysteresis around $T_C$. On the
contrary, the low signal and the lack of thermal hysteresis at
$T^*$ in the indide compound indicates that no magnetic transition
occurs at that temperature.

\subsubsection{\bf Magnetic properties}

The $T_M(Pd)$ dependence was also traced by $M(T)$ measurements.
In Fig.~\ref{F3} we show the inverse of the low temperature
susceptibility measured with $B=0.3$\,T on two representative
samples: Ce$_{1.95}$Pd$_{2.2}$In$_{0.85}$ and
Ce$_{1.95}$Pd$_{2.25}$In$_{0.8}$. The respective paramagnetic
temperatures $\theta_P$ and AF transitions $T_M$ are depicted in
Fig.~\ref{F3} as a function of $Pd$ concentration (right and upper
axes) for all studied alloys, including a FM one with
$Pd(2+y)=1.95$. One can see that $\theta_P(Pd)$ rapidly decrease
with $Pd$ content, extrapolating to zero for $Pd(2+y)\approx
2.43$, whereas $T_M(Pd)$ decreases moderately.

\begin{figure}
\begin{center}
\includegraphics[angle=0, width=0.55 \textwidth] {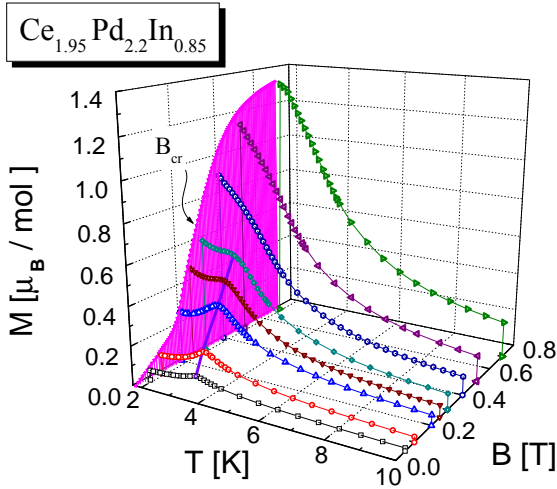}
\end{center}
\caption{(Color online) Thermal dependence of the magnetic
isopedias for different applied fields. At $T=1.8$\,K a $M(B)$ is
included and the continuous line at $T\approx 3$\,K connect the
$M(T_M)$ cusps. $B_{cr}$ indicate the critical field at which the
$T_M$ transition vanishes.} \label{F4}
\end{figure}

The measurement performed up to room temperature on the
Ce$_{1.95}$Pd$_{2.20}$In$_{0.85}$ sample (not shown) allowed to
evaluate the magnetic moments of the ground and excited crystal
field CF levels, and their respective CF splitting. For such a
purpose we have used the simplified formula \cite{PdRh}:

\begin{equation}
M/B =\sum_0^2 \mu_i^2*exp(-\Delta_i/T)/ [(T-T_I)*Z]
\end{equation}

where $\mu_i$ are the effective moments of the ground state (i=0)
and excited CF levels (i=1, 2 respectively). The values obtained
for the ground and first CF excited levels: $\mu_0=1.6\mu_B$ and
$\mu_1=2.4\pm 0.2\mu_B$, with a CF splitting: $\Delta_1=60\pm 5$K.
These values are very similar to those obtained for Ce$_2$Pd$_2$Sn
\cite{Ce2Pd2Sn} and guarantee that only the GS doublet is
responsible for the low temperature properties. As a consequence
of the good fit obtained by applying eq(1) without including any
{\it 4f-conduction band} hybridization effect, one conclude that
this system can be considered as formed by Ce$^{3+}$ ions lattice
with irrelevant Kondo effect (i.e. $T_K < T_M$).

\subsubsection{\bf Field dependencies}

The temperature dependence of magnetization $M(T)$ was studied in
all the samples under different applied fields up to $B=0.7$\,T in
order to obtain the respective magnetic phase diagrams. In
Fig.~\ref{F4} we present a detailed investigation performed on the
Ce$_{1.95}$Pd$_{2.20}$In$_{0.85}$ alloy, selected for having the
composition closer to stoichiometry. Albeit those $M(T)$ isopedias
were measured between $1.8 \leq T \leq 20$\,K, they are only shown
up to 10\,K for clarity. The $M(B)$ isotherm obtained at our
lowest measured temperature ($T=1.8$\,K) is also included in the
$M-B$ plane. These $M(T)$ curves show an AF type behavior up to
about $B=0.18$\,T, turning into a field induced FM polarization at
higher field. As this polarization arises, it overcome the $T_M$
transition defined as the cusp of $M(T)$ (see the continuous curve
around $T=3$\,K in Fig.~\ref{F4}). This indicates that the ShSu
phase is suppressed at a critical field $B_{cr}$ slightly above
$B=0.3$\,T. Notably, $T_M$ itself is slightly affected by magnetic
field because in that alloy it decreases from 3.56K at 50\,mT to
3\,K at 0.3\,T. This indicates that the intensity of the $J$ and
$J'$ couplings are weakly affected by field despite of the degrees
of freedom progressively transferred to the FM phase. This field
dependence for all the studied alloys is included in the magnetic
phase diagrams presented at the end of the discussion section.

\begin{figure}
\begin{center}
\includegraphics[angle=0, width=0.5 \textwidth] {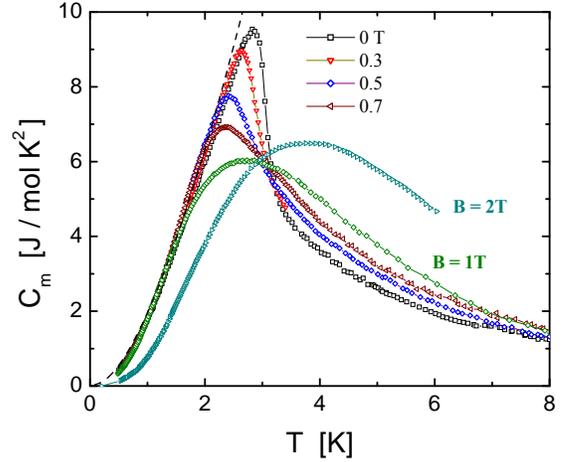}
\end{center}
\caption{(Color online) Magnetic field effect on the specific heat
of Ce$_{1.9}$Pd$_{2.4}$In$_{0.7}$ up to $B = 2$\,T. Dashed curve
is the fit with for $B=0$ using eq.(2).} \label{F5}
\end{figure}

The magnetic field effect on the specific heat $C_m(T)$ was also
studied in all the samples, however more detailed measurements
were carried on the richest $Pd$ sample
Ce$_{1.9}$Pd$_{2.4}$In$_{0.7}$, up to $B=2$\,T as shown in
Fig.~\ref{F5}. Since this alloy lies at the edge of the solid
solution formation, we have chosen it for the field effect
investigation in order to drive $T_M$ to lower temperature and to
trace the disappearance of the $T^*$ shoulder. The temperature of
the $C_M(T)$ maximum (coincident with $T_M$ defined by the cusp in
$M(T)$) progressively decreases with field till it reaches the
value of $T^*$ at $B_{cr}=0.7$\,T. Beyond that value the $C_m(T)$
maximum broadens and start to increase in temperature with a
typical behavior of a FM field-polarized system. Coincidentally,
the $C_m(T_M)$ jump is overcome by the high temperature tail of
$C_m(T>T_M)$, indicating an increase of the magnetic correlations
prior to the transition. Up to $B = 0.5$\,T the low temperature
($T<T^*$) $C_m(T)$ dependence practically coincide, without
affecting substantially the gap of anisotropy nor the $T^{1.5}$
coefficient determined applying eq.(2), see next section. Since
$C_m(T)$ curves crosses each other (at $\approx 3$\,K), one may
conclude that there are degrees of freedom transferred from the
intermediate phase (i.e. $T^* < T < T_M$) to the region where
dimers form (i.e. $T>T_M$). On the contrary, no degrees of freedom
are transferred from $T<T^*$ by effect of the magnetic field. This
indicates that the FM GS is more robust under field than the
intermediate phase between $T_M$ and $T^*$.

\section{Discussion}

The magnetic nature of the GS can be recognized analyzing the
thermal dependence of the specific heat. Below $T^*$, the $C_m(T)$
dependence is properly fitted according to an anisotropic FM
relation dispersion like the observed in Ce$_2$Pd$_2$Sn
\cite{Ce2Pd2Sn}:

\begin{equation}
C_m(T<T^*) = \gamma T+ A T^{3/2} \exp(-\Delta_M/T)
\end{equation}

where $\gamma$ is the Sommerfeld coefficient and $\Delta_M$ a gap
in the magnon spectrum which rises from zero for $Pd(2+y)= 2.2$ up
to 0.4\,K for the $Pd(2+y)= 2.4$. Such an increase of $\Delta_M$
indicates a increasing anisotropy driven by the fact that the
Ce-Ce inter-plane distance expands from $3.963\,\AA$ to
$4.017\,\AA$ \cite{Mauro00}. The $\gamma$ coefficient extracted
for these Pd-rich AF alloys is comparable to that of the $La$
isotypic compound in agreement with the expected Ce$^{3+}$
character of the magnetic ions and the consequent negligible Kondo
effect.

\begin{figure}
\begin{center}
\includegraphics[angle=0, width=0.55 \textwidth] {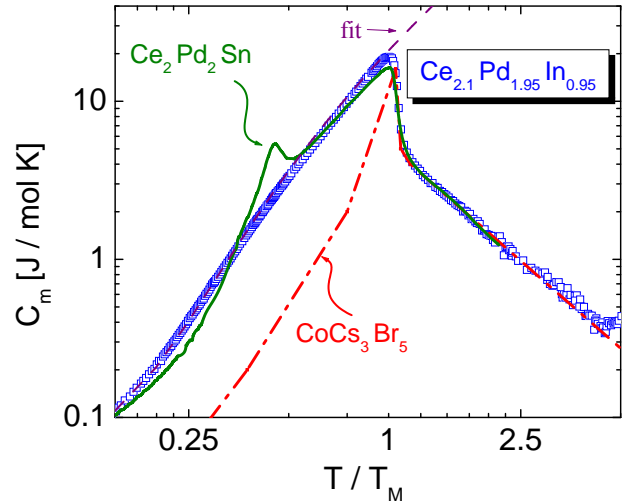}
\end{center}
\caption{(Color online) Comparison of the specific heat magnetic
contribution of two Ce$_2$Pd$_2$X (X=In and Sn) compounds with the
model compound for 2D simple square lattice (dashed dot line), all
normalized to their respective ordering temperatures. Dashed curve
is the fit to Ce$_{2.1}$Pd$_{1.95}$In$_{0.95}$ using eq(1).}
\label{F6}
\end{figure}

For the case of the studied Ce-rich FM alloy
Ce$_{2.1}$Pd$_{1.95}$In$_{0.95}$, a similar thermal dependence is
observed, with $C_m = 0.14 + 4.3 T^{3/2} \exp(-2.8/T)$ up to
$T_C$, see Fig.~\ref{F6}. A FM type dispersion relation is
certainly expected because of the spontaneous magnetization
observed in $M(T)$ below $T_C = 4$\,K. However, the large $\gamma
= 0.14$\,J/mol K$^2$ is somehow unexpected from the 3$^+$
character of the Ce atoms and in comparison with that of the AF
alloys. This enhanced contribution can be explained by a closer
inspection to the interatomic Ce-Ce spacing on the Ce-rich
FM-branch provided in Ref. \cite{Mauro00}. Although Ce-Ce spacing
on plane slightly increases following the 'a' lattice parameter,
the situation around the extra Ce atom loci on the $z= 0$ plane
(at the '$2a$' Wyckoff position) is quite different. Due to the
larger atomic radius of Ce atoms with respect to $In$ ones, the
substituted atoms support a significant 'crystal pressure' which
reduces the Ce-Ce spacing below the magnetic limit. As a
consequence, hybridization effects may arise between those
substituent atoms and their Ce-$nn$ on the '$4h$' position in the
$z=0$ plane. In such a case, those atoms may contribute to the
specific heat as Kondo impurities, increasing the $\gamma$
coefficient.

Concerning the low temperature properties presented in
Fig.~\ref{F3}, there is an apparent contradiction between positive
value of $\theta_P$ expected for a FM system and the AF like cusp
in temperature dependence of the susceptibility. Like in
Ce$_2$Pd$_2$Sn, this fact indicates that two contributions are
competing at that temperature. Both mechanisms are intrinsic to
the formation of a ShSu lattice because the FM $J$ coupling is
responsible for the dimers formation above $T_M$ and the AF $J'$
couples those dimers to build up the ShSu network below $T_M$.

\subsection{Proving dimers formation}

Since the objective of this work is to investigate the effect of
electronic (holes) variation on the ShSu lattice stability, the
first issue is to ascertain whether dimers actually form in these
alloys like in Ce$_2$Pd$_2$Sn. The second is to check whether the
magnetic phase below $T_M$ keeps having the characteristics
assigned to ShSu lattices once Sn [5s$^2$5p$^2$] is replaced by In
[5s$^2$5p$^1$] and the number of holes is further increased by
$Pd$ concentration.

\begin{figure}
\begin{center}
\includegraphics[angle=0, width=0.5 \textwidth] {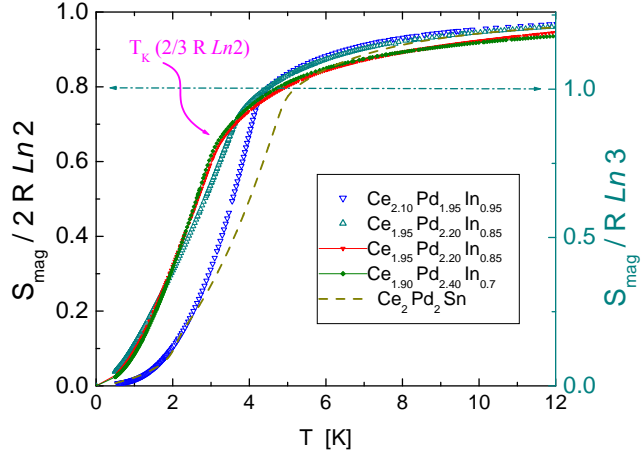}
\end{center}
\caption{(Color online) Comparison of the thermal increase of the
entropy between indide and stanide compounds. $T_K \simeq 2/3
R\ln2$ indicates the curve from which the Kondo temperature is
evaluated.} \label{F7}
\end{figure}

Following the same procedure proposed in Ref.\cite{Ce2Pd2Sn} we
have evaluated the magnetic entropy gain $S_{mag}(T)$ from the
measured specific heat as $S_{mag}=\int C_m/T dT$, and included
those results in Fig.~\ref{F7} for different $Pd$ concentrations,
including one Ce-rich FM alloy. One can see that for the latter
compound the 'per Ce atom' entropy at $T= T_M$ is $S_{mag}=0.8 R
Ln 2$, which coincides with the molar entropy of $R Ln 3$. Like
for stoichiometric Ce$_2$Pd$_2$Sn \cite{Ce2Pd2Sn}, included in the
figure as reference, the formation of FM dimers is inferred. The
remaining entropy is collected above the transition since the
onset of magnetic correlations occurs at $T\leq 20$\,K as
indicated by the $C_m(T>T_M$) tail in Fig.~\ref{F6}. Notice that
for the Ce$_{2.1}$Pd$_{1.95}$In$_{0.95}$ compound $C_m(T>T_M$)
nicely coincides with that of the model compound for the 2D-square
lattice CoCs$_2$Br$_5$ \cite{joung} and the parent compound
Ce$_2$Pd$_2$Sn.

For the alloys belonging to the AF branch, the entropy at $T_M$
decreases progressively and the criterium to compare $S_{mag}=0.8
R Ln 2$ with $R Ln 3$ can be applied only to the sample close to
the stoichiometry, i.e. $Pd_{2+y}=2.2$. For higher $Pd$
concentrations the entropy evaluation has to be normalized to the
actual number of Ce atoms not affected by the proximity of a
double $Pd$ occupation in sites '4e'. This re-normalization may
include the $Pd_{2+y}=2.25$ sample into those where Ce-Ce dimers
are formed, but it does not apply satisfactorily to the samples
with $Pd_{2+y}=2.35$ and $2.4$. Nevertheless, the increase of
double occupied '4e' sites progressively destroys the lattice
character of the ShSu network.

The value of the Kondo temperature for this system can be
evaluated also from the $S_{mag}(T)$ results shown in
Fig.~\ref{F7} by applying the Desgranges-Schotte \cite{Desgr}
criterion of $S_{mag}(T = T_K) \simeq 2/3 R\ln2$ for a non-ordered
system. As shown in Fig.~\ref{F7} $T_K \approx 3$\,K, which is
similar to the value found for the parent compound Ce$_2$Pd$_2$Sn
\cite{Ce2Pd2Sn}. This is one of the lowest values found among the
Ce compounds and confirms the $Ce^{3+}$ character of this ion.

\subsection{Proving SSL symptoms}

\begin{figure}
\begin{center}
\includegraphics[angle=0, width=0.52 \textwidth] {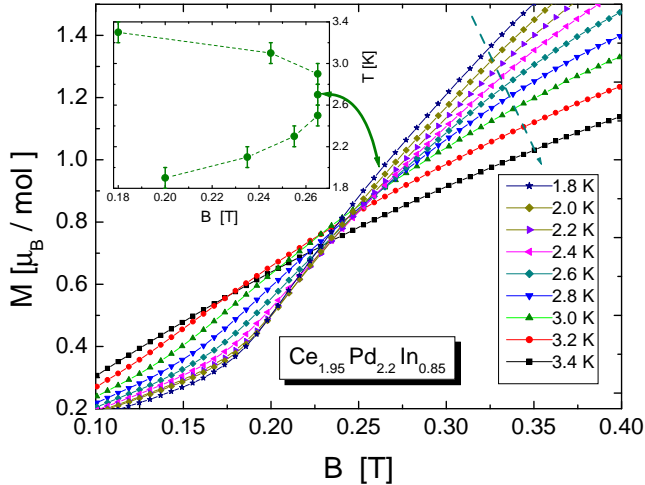}
\end{center}
\caption{(Color online) Detail of the M vs B isotherms of the
Ce$_{1.95}$Pd$_{2.2}$In$_{0.85}$ sample in the region where the
crossing occurs. In the inset the temperature and field values of
the crossing points are depicted.} \label{F8}
\end{figure}

As mentioned in the introduction, one of the characteristic
features observed in the exemplary compounds exhibiting a ShSu
lattice is the crossing of the $M(B)$ isotherms at low
temperatures. In order to search for this effect in the family of
compound under study we have chosen the
Ce$_{1.95}$Pd$_{2.2}$In$_{0.85}$ alloy because it shows the more
pronounced change of slope of $M(B)$ within our range of
measurement. In Fig.~\ref{F8} we show the detailed study of those
isotherms in the region of temperature between $1.8 \leq T \leq
3.4$\,K. Within the experimental dispersion, one can see that such
a crossing occurs between $2.4 \leq T \leq 3$\,K at a field $B =
0.26 \pm 0.05$\,T, where the magnetic moment is $M = 0.89 \pm 0.01
\mu_B$/mol, see the inset in Fig.~\ref{F8}. For higher $Pd$
concentration (c.f. $Pd(2+y)=2.25$) the crossing occurs at higher
field and slightly lower temperature, albeit the slope of the
isotherms is not so pronounced and the effect practically vanishes
for $Pd(2+y)= 2.35$.

Some thermodynamical consequences can be deduced from the
existence of such a crossing. Since within that window of
temperature the magnetization does not change, then $\partial
M/\partial T\mid_B = 0$, and from Maxwell relations also $\partial
S/\partial B\mid_T = 0$ is deduced. This implies the existence of
an 'isosbestic' point \cite{Voll}, which should be reflected in
the corresponding thermodynamic parameters like $C_m/T (T,B)$.
Effectively specific heat measured at $B=0$ and $0.5$\,T (not
shown) cross each other at $\approx 2.9K$.

\begin{figure}
\begin{center}
\includegraphics[angle=0, width=0.55 \textwidth] {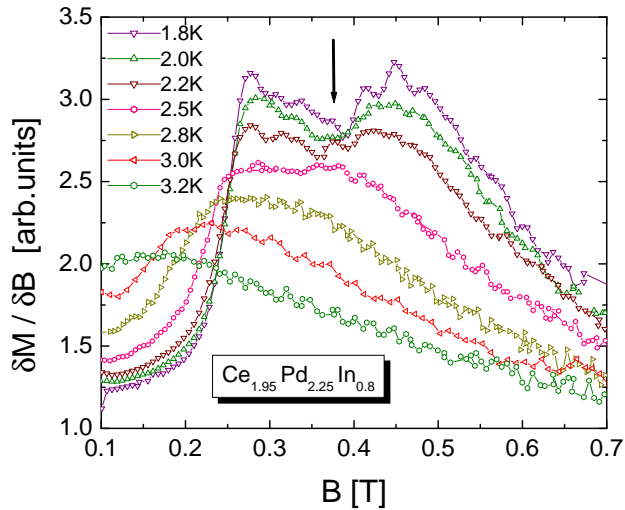}
\end{center}
\caption{(Color online) Field dependence of the M(B) derivative to
show the a relative minimum related with an incipient step in $M
vs B$ isotherms, The arrow indicates the mínimum of that
derivative which is related with the incipient step.} \label{F9}
\end{figure}

The other characteristic of the SSL is the appearance of quantized
$M(B)$ steps. From the measurements of the $M(B)$ isotherms no
discontinuity was observed probably due to the poly-crystalline
nature of the samples. However, the analysis of their derivatives
$\partial M/\partial B\mid_T$ allows to detect an incipient
structure as a function of field as it is shown in Fig.~\ref{F9}.
Notice that for that figure we have selected the
Ce$_{1.95}$Pd$_{2.25}$In$_{0.8}$ alloy because this effect occurs
at the lower limit of temperature $T=1.8$\,K reached by our
magnetometer. Thus, we took profit from the fact that this effect
occurs at higher temperature as $Pd$ content increases and
measured the $Pd(2+y)= 2.25$ sample where at least three isotherms
show the mentioned structure in their $\partial M/\partial B(B)$
dependence. Further increase of $Pd$, i.e. for $Pd(2+y)= 2.35$ and
$2.4$, strongly weakens the $M(B)$ curvature smearing out the
effect. As shown in Fig.~\ref{F9} the minimum related with the
incipient step occurs at $B\approx 0.4$\,T, where $M(B) =
0.4\,\mu_{B}$ is $\approx 1/4 M_{sat}$ since $M_{sat}\approx
1.3\,\mu_{B}$ \cite{Mauro00}.

Noteworthy is the comparison between the different fields at which
the $M(B)$ steps occur and the nearly constant temperature of the
crossing of $M$ isotherms for the different mentioned systems.
Whereas in SrCu(BO)the plateaux occur at very high field $B\approx
20$\,T, in metallic RB$_4$ it occurs around $4 < B < 8$\,T
depending on the crystallographic directions, and at much lower
$B$ (a fraction of Tesla) in R$_2$T$_2$X compounds. In spite of
that, the crossing of $M(B)$ isotherms always occurs at low
temperature ($T<4$\,K). This significant difference between
magnetic and thermal scales support the role of the 'quantized
magnetization' constrain, making relevant the fact that these
effects are related to fractional $M/M_{sat}$ values. Since in
SrCu$_2$(BO$_3$)$_2$ a super-exchange mechanism drives the
magnetic interactions, high applied field is required to reverse
the magnetic nature of the GS in the singlet-triplet level
spectrum of the dimers. On the contrary, for RKKY mediated
magnetic interactions such a singlet-triplet modification is
expected to be much weaker, and consequently lower field would
induce the 'quantized magnetization jumps'. On the other hand, the
crossing effect depends on relative thermal population of the
levels. Once the field driven values for the appropriated
$M/M_{sat}$ values is reached, field independent thermal
excitations produce the 'crossing effect'. In other words, this $M
vs. T$ effect cannot the scaled as $M vs. B/T$.

\subsection{Phase Diagram}

\begin{figure}
\begin{center}
\includegraphics[angle=0, width=0.5 \textwidth] {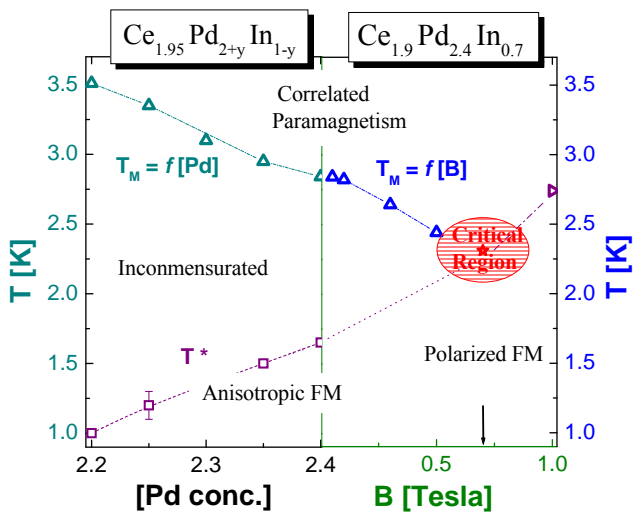}
\end{center}
\caption{(Color online) Magnetic Phase diagram as a function of a)
$Pd$ concentration and b) field dependence for the $Pd_{2.4}$
sample. The arrow indicates the critical field at which the
minimum value of $T_{max}(C_P)$ is observed.} \label{F10}
\end{figure}

The magnetic phase diagram as a function of $Pd$ concentration is
shown in Fig.~\ref{F10}a (left side) with $T_M(Pd)$ defined as the
temperature of the maximum value of $C_m(T)$ and $T^*(Pd)$ as the
temperature of the maximum curvature of $C_m(T)/T$ defined as a
kink for $Pd(2+y)= 2.20$ and $2.25$, and as a 'shoulder' for
$Pd(2+y)= 2.35$ and $2.40$ in Fig.~\ref{F1}.

This $y(Pd)$ dependent phase diagram is complemented by the field
dependence of the $Pd(2+y)= 2.4$ sample after an appropriated
scaling between $Pd$ concentration and magnetic field on the right
side of Fig.~\ref{F10}b. Although the second order character of
the magnetic transition smears progressively out in the $Pd(2+y)=
2.4$ sample under applied field, the maximum of $C_m(T,B)$ allows
to extend the upper phase boundary up to a magnetic critical point
at $T_{cr} = 2.3$\,K and $B_{cr} = 0.7$\,T.

\begin{figure}
\begin{center}
\includegraphics[angle=0, width=0.55 \textwidth] {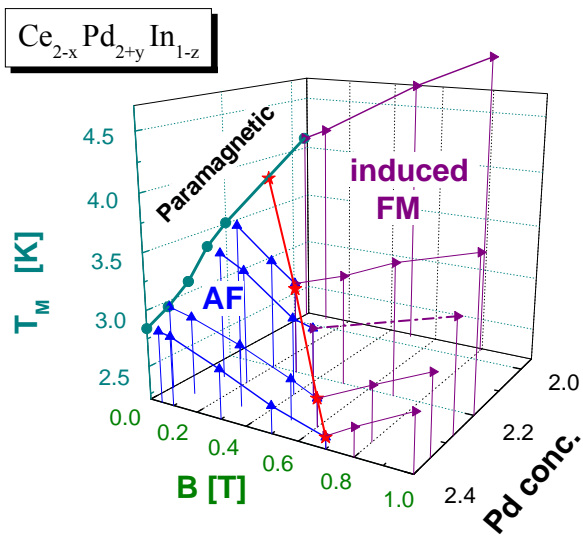}
\end{center}
\caption{(Color online) 3D representation of the magnetic phase
diagram as a function of $Pd$ concentration and magnetic field.}
\label{F11}
\end{figure}

A 3D representation of the magnetic phase diagram is presented in
Fig.~\ref{F11} as a function of both control parameter: field and
$Pd$ concentration, in order to compare the $B$ dependence of all
the AF alloys. In this diagram also one of the Ce-rich FM samples
(c.f. $Pd(2+y) = 1.95$) is included with the purpose to test the
$Pd$ concentration (i.e. the 'hole' concentration) as unique
parameter able to encompassing the alloys studied in this work,
even beyond the gap of miscibility. No further extrapolation on
the $Pd$ dependence can be done within the Ce-rich FM side because
in that 'branch' of the phase diagram the free concentration
parameter is the $Ce$ composition, whilst the $Pd$ remains
constant \cite{Mauro00}.

The minima of $T_M$ as a function of field observed in the AF
alloys indicate that the critical field $B_{cr}(Pd)$, where the
polarized FM behavior takes over, increases with the $Pd$ content.
As expected, the magnetic boundary for $Pd(2+y) = 2.4$ coincides
with the right side of Fig.~\ref{F10}. Notably, the extrapolation
of  $B_{cr}(y) \to 0$ occurs at $Pd(2+y) = 2.1$, which lies within
the gap between AF and FM branches.

\section{Conclusions}

The magnetic properties of the Ce$_{2-x}$Pd$_{2+y}$In$_{1-z}$
family of compounds can be described as a function of a unique
parameter: the decrease of the electronic concentration, and
consequently their Fermi energy, as the number of 'holes' increase
with $Pd$ concentration.

Their ground state shows a common ferromagnetic character in their
magnon dispersion extracted from the specific heat at low
temperature. This is confirmed by magnetization measurements.
However the transition from their respective paramagnetic states
is different depending on the relative $Ce/Pd$ concentration. The
rich $Ce$ sample shows a typical FM transition. On the contrary,
the rich $Pd$ ones show a transition with AF characteristics which
become FM after a characteristic temperature $T^*$.

Since the value extracted for the Kondo temperature is extremely
low, the magnetic $4f$ states of the $Ce$ ions are strongly
localized all along the series. Nevertheless, in the $Ce$ rich
sample, the $Ce$ atoms substituting $In$ seem to behave
differently since their available volume is considerably smaller.

The conditions for a lattice of dimer formation is realized only
in the alloys lying closer to stoichiometry, i.e. in the $Pd(2+y)=
2.20$ and $2.25$ ones. Higher $Pd$ concentration distorts the
Ce-lattice inhibiting the formation of a well defined network of
those dimers. This reveals the instability of the ShSu lattice
with in front of atomic disorder.

Concerning the ShSu phase symptoms in the $Pd(2+y)= 2.20$ and
$2.25$ alloys, the $M(B)$ isotherms crossing is observed in the
alloys of that AF branch. However, only an incipient modulation in
the $M(B)$ dependence is seen in its derivative. The
ploy-crystalline character of this system may explain the weakness
of this effect together with its intermetallic nature since the
electrons in the conduction band (responsible for the RKKY
interaction) are strongly delocalized.

An interesting complement to this research can be to drive the
weakening of the $Ce^3+$ moments by chemical pressure.

\section*{Acknowledgments}
This work was partially supported by a CNR (Italy) and CONICET
(Argentina) cooperation program, PICTP-2007-0812 and Secyt-UNC
06/C256 projects.

\end{document}